\documentstyle[pra,aps,epsfig,eqsecnum,amssymb,amsmath,amsthm]{revtex}
\tighten
\begin{document}
\clearpage
\preprint{}
\draft

\title{Global-fidelity limits of state-dependent cloning of mixed
states}

\author{ Alexey E. Rastegin }
\address{Department of Theoretical Physics, Irkutsk State University,
Gagarin Bv. 20, Irkutsk 664003, Russia}

\maketitle
\begin{abstract}
By relevant modifications, the known global-fidelity limits of
state-dependent cloning are extended to mixed quantum states.
We assume that the ancilla contains some {\it a priori} information
about the input state. As it is shown, the obtained results
contribute to the stronger no-cloning theorem. An attainability of
presented limits is discussed.
\end{abstract}

\pacs{03.67.-a, 03.65.Ta}

\pagenumbering{arabic}
\setcounter{page}{1}

\protect\section{Introduction}

It is well known that a quantum state seeks to escape the observer.
One of the manifestations of this sensitivity is expressed by the
uncertainty relation. Other striking disclosure is made by
the no-cloning theorem of Wootters, Zurek \cite{wootters} and Dieks
\cite{dieks}.  In effect, an arbitrary unknown pure state cannot be
cloned. With relevant modifications, this result is generalized and
extended to mixed quantum states \cite{barnum}. That is,
noncommuting mixed states cannot be broadcast. Pati and Braunstein
formulated the no-deleting principle \cite{pati}, which is
complementary to the no-cloning theorem. In Ref.
\cite{jozsa1} the stronger no-cloning theorem was established. Let
$\,\{|s_1\rangle,|s_2\rangle\}\,$ be any pair of nonorthogonal pure
states and $\,\{\Upsilon_1,\Upsilon_2\}\,$ be any pair of mixed
states.  According to the stronger no-cloning theorem, there is a
physical operation $\:|s_j\rangle\otimes\Upsilon_j \longmapsto
|s_j\rangle |s_j\rangle\:$ if and only if there is a physical
operation $\:\Upsilon_j \longmapsto |s_j\rangle\:$. In other words,
the full information of the clone must {\it a priori} be provided in
the ancilla state $\Upsilon_j$ alone \cite{jozsa1}.

Thus, it is impossible to copy an arbitrary unknown quantum state
perfectly. However, nothing prevents us from finding a close
approach to the ideality. The approximate quantum copying
was originally considered by
Bu\v{z}ek and Hillery \cite{buzek}. In addition, they examined
approximate cloning machines operating on prescribed two
nonorthogonal pure states \cite{hillery}. In Ref. \cite{brass} such
devices were called "state-dependent cloners". It is clear that
evaluation of an accuracy of cloning is dependent on the notion of
"closeness" to ideality. The authors of Ref. \cite{brass} introduced
the notions of "global fidelity" and "local fidelity". The
state-dependent cloning was studied by optimizing both the global
fidelity and the local fidelity. Chefles and Barnett generalized
the notion of the global fidelity to the case, where
the states to be cloned have arbitrary {\it a priori}
probabilities \cite{chefles}. They obtained the least upper bound on
the global fidelity for cloning of two pure states with arbitrary
{\it a priori} probabilities. Ref. \cite{macchi} considered
state-dependent $N\to L$ cloning with respect to both "global
fidelity" criterion and "local fidelity" criterion. The other
category of cloners contains universal cloning machines which copy
arbitrary state equally well. First such example was given by
Bu\v{z}ek and Hillery \cite{buzek}. Refs. \cite{brass,gisin}
constructed the universal qubit cloner that maximizes the local
fidelity. Analogous problem for multi-level quantum system was solved
in Refs. \cite{werner,keyl}.

All the above results examine the pure-state cloning. Ref.
\cite{cirac} introduced the single qubit purification procedure that
was used in extending of the acceptable input for the optimal cloners
to mixed quantum states. However, the described in Ref. \cite{cirac}
scenario is not equivalent to the standard statement of cloning
problem.

Ref. \cite{rast1} presented a new approach to the problem of
state-dependent cloning. It examined the relative error which is
complementary to the criterion of the global fidelity. The method of
Ref. \cite{rast1} lively uses the notion of the angle between two
vectors. Ref. \cite{rast2} extended the notion of the angle to mixed
quantum states, that allows to widen the global-fidelity limit of
state-dependent cloning deduced in Ref. \cite{brass} to the
mixed-state cloning. For this extension the squared modulus of the
inner product should be replaced by the fidelity function of states
to be cloned. Ref. \cite{rast3} obtained the lower bound on the
relative error of mixed-state cloning and related physical
operations, in which the ancilla contains some {\it a priori}
information about the input state. The authors of Ref. \cite{guo}
modified the method of Ref. \cite{rast1} and considered the
state-dependent cloning of set which contains $n>2$ pure states. They
deduced the upper bound on the global fidelity for the case of equal
{\it a priori} probabilities of states.

In this paper, we extend the known bounds on the global fidelity of
state-dependent cloning to mixed quantum states. This is
attained by natural development of the method of Ref. \cite{rast2}.
The bounds arise from the unitarity of quantum mechanical
transformations and that the fidelity function cannot decrease under
the operation of partial trace. We consider the case, in which the
ancilla contains some {\it a priori} information about the state to
be cloned. The relationship between obtained limits and the stronger
no-cloning theorem is discussed.

\protect\section{Statement of the problem}

Let us start with a precise formulation of the problem. A register $A$
is composed of $N$ systems, each having an $d$-dimensional Hilbert
space ${\cal{H}}={\mathbb{C}}{\,}^d$ ($d>1$). Initially, every of
these $N$ systems is with probability $p_j$ prepared in one and the
same state $\rho_j$ from a set
$\,{\mathfrak{A}}=\{\rho_1,\ldots,\rho_n\}\,$. We want making $L>N$
copies of the given $N$ systems. In order to attain this we use the
ancilla which contains some {\it a priori} information about the
input state $\rho_j$. That is, the ancilla is initially prepared in
state $\Upsilon_j$ from a set
$\,{\mathfrak{S}}=\{\Upsilon_1,\ldots,\Upsilon_n\}\,$ indexed by the
same labels. We will mean that the ancilla $BE$ is combined of extra
register $B$ and environment $E$. Of course, the extra register $B$
contains $M=L-N$ additional systems, each is to receive the clone of
$\rho_j$. Thus, the final state of two registers is described by
\begin{equation}
\widetilde{\rho\,}_j={\rm Tr}_{E}
\left({\rm U}(\rho_j^{\otimes N}\otimes\Upsilon_j)
{\rm U}^{\dagger}\right) \!\ ,
\label{fistat}
\end{equation}
which is partial trace over environment space.

In order to estimate an accuracy of cloning we must of course make
use of some quantitative measure of distinguishability for mixed
states. This need is met by the fidelity function \cite{jozsa}.
There are several ways to the definition of fidelity. The general
expressions is given by
\begin{equation}
F(\chi,\omega)=
\left\{{\rm Tr}
\! \left[\bigl(\sqrt{\chi}\,\omega\sqrt{\chi}\,\bigr)^{1/2}\right]
\right\}^2 \!\ ,
\label{uhldef}
\end{equation}
where the trace is taken over the same space on which the density
operators $\chi$ and $\omega$ are considered. The right-hand side of
Eq. (\ref{uhldef}) is the transition probability between mixed states
$\chi$ and $\omega$ introduced by Uhlmann \cite{uhlmann}. Jozsa
proposed the definition in terms of purifications \cite{jozsa}; it
provides a kind of physical interpretation of Eq. (\ref{uhldef}).
Note that the concept of purifications of mixed states is a natural
development of the "decoherence" point of view.
According to this viewpoint \cite{zurek}, any mixed state is
really describing the reduced state of a subsystem entangled with a
larger system. The total system is always being in a pure state
described by vector in a Hilbert space. As it is shown in
Ref. \cite{jozsa}, the fidelity function is an analogue of the
squared modulus of the inner product for pure states  The fidelity
can also be defined by means of generalized measurements
\cite{barnum}. (Note that Refs.  \cite{barnum,uhlmann1} define
fidelity to be the square root of the right-hand side of Eq.
(\ref{uhldef}).) In Ref. \cite{rast2} we parametrized the fidelity
function by
\begin{equation}
F(\chi,\omega)=
\cos^2\!\Delta(\chi,\omega) \!\ ,
\label{link}
\end{equation}
where the angle $\Delta(\chi,\omega)$ ranges between $0$ and $\pi/2$.
Using the concept of purifications, we extended the triangle
inequality to the case of mixed states. That is \cite{rast2},
\begin{equation}
\Delta(\chi,\omega)
\leq\Delta(\chi,\rho) + \Delta(\omega,\rho)
\!\ .
\label{choose}
\end{equation}
Other properties of the angle between two mixed states are listed in
Ref. \cite{rast2}. Note that $\,\sin\Delta(\chi,\omega)\,$ provides a
reasonable measure of closeness for mixed states $\chi$ and $\omega$.
For any (generalized) measurement \cite{rast3},
\begin{equation}
\bigl|\,p(a|\chi) - p(a|\omega)\,\bigr|
\leq \sin\Delta(\chi,\omega)
\!\ ,
\label{inpr}
\end{equation}
where $p(a|\chi)$ and $p(a|\omega)$ are the probabilities of outcome
$a$ generated by $\chi$ and $\omega$ respectively.

Thus, the fidelity function generalizes the squared modulus of the
inner product. Therefore, it is natural to define the global fidelity
of mixed-state cloning as
\begin{equation}
F_G=\sum_{1\leq j\leq n} p_j
\, F(\widetilde{\rho\,}_j,\rho_j^{\otimes L}) \!\ .
\label{global}
\end{equation}
The definition by Eq. (\ref{global}) uses  the state
$\,\rho_j^{\otimes L}\,$ as ideal output. So we consider the cloning
just. Recall that the cloning is a special strong form of
broadcasting \cite{barnum}.  The broadcasting is most general type of
quantum copying. By broadcasting Ref.  \cite{barnum} means that the
marginal density operator of each of the separate systems is the same
as the input state to be broadcast. The cloning case is specified by
choice of state $\,\rho_j^{\otimes L}\,$ as perfect. Replacing the
squared modulus of the inner product by the fidelity function, the
present definition extends the definition of Ref. \cite{chefles} to
mixed quantum states.

We are interested in a nontrivial upper bound on the global fidelity
defined by Eq. (\ref{global}). Our approach to obtaining the limits
employs triangle inequalities and general properties of the fidelity
function. Following the method of Ref. \cite{rast2}, we shall derive
the angle relation from which bound on the global fidelity is simply
obtained. In order to be rid of  bulky expressions we shall use the
notation
\begin{align}
\Delta^{(L)}_{jk} &=\Delta(\rho_j^{\otimes L},\rho_k^{\otimes L})
\!\, \\
\delta_j &=\Delta(\widetilde{\rho\,}_j,\rho_j^{\otimes L}) \!\ .
\end{align}
We also introduce the angle
\begin{equation}
\alpha_{jk}=\arccos\left[
F(\rho_j^{\otimes N},\rho_k^{\otimes N})\, F(\Upsilon_j,\Upsilon_k)
\right]^{1/2} \!\ ,
\label{alpha1}
\end{equation}
which lies in the interval $[0;\pi/2]$.

\protect\section{Limit for two-state set}

In this section we establish the limit of state-dependent cloning of
two-state set $\,{\mathfrak{A}}=\{\rho_1,\rho_2\}\,$. The initial
state of ancilla is $\Upsilon_1$ or $\Upsilon_2$ according to the
input state which is $\rho_1$ or $\rho_2$. We shall restrict our
consideration to the case in which
\begin{equation}
F(\rho_1^{\otimes M},\rho_2^{\otimes M})
<F(\Upsilon_1,\Upsilon_2)
\!\ .
\label{restr}
\end{equation}
The motivation for this restriction consists in the following. As it
is shown in Appendix A, if Eq. (\ref{restr}) is not satisfied then
there are states sufficient for perfect cloning. That is, there are
states $\Upsilon_1$ and $\Upsilon_2$ such that
$$
\rho_j^{\otimes M}={\rm Tr}_{E}\Upsilon_j
$$
for $j=1,2$. Here we can only point to trivial bound $F_G\leq1$. So
we presuppose that Eq. (\ref{restr}) is valid.  As result, we have
\begin{equation}
\alpha_{12}<\Delta^{(L)}_{12} \!\ .
\label{alpdel}
\end{equation}
The desired limit is established by the following theorem.

{\bf Theorem 1} {\it The global fidelity $F_G$ of $N\to L$ cloning for
set $\,{\mathfrak{A}}=\{\rho_1,\rho_2\}\,$ is limited
above by value}
$$
\frac{1}{2}
\left\{1+\left[
1-4p_1p_2\sin^2(\Delta^{(L)}_{12}-\alpha_{12})
\right]^{1/2}\right\} \!\ .
$$

{\bf Proof}
Applying Eq. (\ref{choose}) twice, we get
\begin{equation}
\Delta^{(L)}_{12}\leq\delta_1+\delta_2+
\Delta(\widetilde{\rho\,}_1,\widetilde{\rho\,}_2)
\!\ . \label{tilde1}
\end{equation}
Recall that the fidelity function is multiplicative and preserved by
unitary evolution \cite{jozsa}. Therefore,
$$
F(\rho_1^{\otimes N},\rho_2^{\otimes N})\, F(\Upsilon_1,\Upsilon_2)
=F\bigl(
\rho_1^{\otimes N}\otimes\Upsilon_1,\rho_2^{\otimes N}\otimes\Upsilon_2
\bigr)=F\Bigl({\rm U}(\rho_1^{\otimes N}\otimes\Upsilon_1)
{\rm U}^{\dagger},{\rm U}(\rho_2^{\otimes N}\otimes\Upsilon_2)
{\rm U}^{\dagger}\Bigr) \!\ .
$$
Because the fidelity cannot decrease under the operation of
partial trace \cite{barnum},
$$
F(\rho_1^{\otimes N},\rho_2^{\otimes N})\, F(\Upsilon_1,\Upsilon_2)\leq
F(\widetilde{\rho\,}_1,\widetilde{\rho\,}_2) \!\ ,
$$
whence we have
\begin{equation}
\alpha_{12}\geq
\Delta(\widetilde{\rho\,}_1,\widetilde{\rho\,}_2)
\!\ .
\label{alpha2}
\end{equation}
Eqs. (\ref{tilde1}) and (\ref{alpha2}) provide
\begin{equation}
\delta_1+\delta_2\geq
\Delta^{(L)}_{12}-\alpha_{12}
\!\ . \label{star1}
\end{equation}
By Eq. (\ref{alpdel}), the right-hand side of Eq. (\ref{star1})
ranges between 0 and $\pi/2$. (Note that if our presupposition given
by Eq. (\ref{restr}) is broken then the right-hand side of Eq.
(\ref{star1}) is nonpositive and Eq. (\ref{star1}) becomes empty.)
According to the definition of the global fidelity,
\begin{equation}
F_G=p_1\cos^2\!\delta_1+p_2\cos^2\!\delta_2 \!\ .
\label{star2}
\end{equation}
We want to maximize the right-hand side of Eq. (\ref{star2}) with the
constraints (\ref{star1}), $\,0\leq\delta_1\leq\pi/2\,$ and
$\,0\leq\delta_2\leq\pi/2\,$. This problem is considered in Appendix
B, the result is formulated as Lemma. Performing the relevant
substitutions, we obtain the statement of Theorem 1. $\blacksquare$

In the case of $1\to2$ cloning of equiprobable states without
{\it a priori} information, the bound given by Theorem 1 is reduced
to the upper bound deduced in Ref. \cite{rast2}. If all the states of
set ${\mathfrak{A}}$ are pure, that is $\,\rho_j=|s_j\rangle\langle
s_j|\,$ for $j=1,2$, and the ancilla contains no {\it a priori}
information, that is $\,\Upsilon_1=\Upsilon_2\,$, then we have
\begin{align*}
\cos\Delta^{(L)}_{12} &=
\bigl|\langle s_1^{\otimes L}|s_2^{\otimes L}\rangle\bigr| \!\ , \\
\cos\alpha^{(N)}_{12} &=
\bigl|\langle s_1^{\otimes N}|s_2^{\otimes N}\rangle\bigr| \!\ .
\end{align*}
In this special case the limit given by Theorem 1 is reduced to the
limit obtained in Ref. \cite{chefles}. Thus, Theorem 1 provides
the extension of the preceding result in two significances. In the
first place, it extends the known limit to the case of mixed
states. In the second place, it takes into account that the ancilla
state can contain {\it a priori} information about the state to be
cloned.

The limit by Theorem 1 is a decreasing function of $p_1p_2$ and
increases as the {\it a priori} probabilities differ. For pure-state
cloning this fact was shown in Ref. \cite{chefles}. So we are rather
interested in dependence of the limit on the parameter
$\,F(\Upsilon_1,\Upsilon_2)\,$. This parameter marks the top amount
of the additional information, which can beforehand be contained in
the ancilla. The more $\,F(\Upsilon_1,\Upsilon_2)\,$ the less the
given amount. Let $\,F(\Upsilon_1,\Upsilon_2)\,$ be variable between
$\,F(\rho_1^{\otimes M},\rho_2^{\otimes M})\,$ and 1 and let the rest
of parameters be fixed. Then we have
$$
\Delta^{(N)}_{12}\leq\alpha_{12}\leq\Delta^{(L)}_{12} \!\ .
$$
In this range the limit by Theorem 1 is an increasing function of
$\alpha_{12}$. In line with Eq. (\ref{alpha1}), the angle
$\alpha_{12}$ decreases as $\,F(\Upsilon_1,\Upsilon_2)\,$
increases. Therefore, the limit by Theorem 1 decreases as
$\,F(\Upsilon_1,\Upsilon_2)\,$ increases. For
$\,F(\Upsilon_1,\Upsilon_2)=F(\rho_1^{\otimes M},\rho_2^{\otimes
M})\,$ the perfect cloning can be attained. In harmony with this, the
limit by Theorem 1 is equal to 1. For example, the equality $F_G=1$
is reached by the ancilla state $\,\Upsilon_j=\rho_j^{\otimes M}\,$,
where $j=1,2$. Here the full information needed for the ideal cloning
is already provided in the ancilla alone. On the contrary, in the
standard cloning there is no {\it a priori} information, i.e.
$\,\Upsilon_1=\Upsilon_2\,$ and $\,F(\Upsilon_1,\Upsilon_2)=1\,$.
Then the limit by Theorem 1 reaches its minimum as a function of
$\,F(\Upsilon_1,\Upsilon_2)\,$. On the whole, these conclusions seem
plausible and contribute to the stronger no-cloning theorem.

For $L$ going to infinity we have $\,\Delta^{(L)}_{12}=\pi/2\,$
(except $\rho_1=\rho_2$) and
\begin{equation*}
\sin^2(\Delta^{(L)}_{12}-\alpha_{12})=
F(\rho_1^{\otimes N},\rho_2^{\otimes N})\, F(\Upsilon_1,\Upsilon_2)
\!\ .
\end{equation*}
Then the global-fidelity limit becomes
$$
F_G\leq\frac{1}{2}
\left\{1+\left[
1-4p_1p_2F(\rho_1^{\otimes N},\rho_2^{\otimes N}) F(\Upsilon_1,\Upsilon_2)
\right]^{1/2}\right\} \!\ .
$$
In the special case of pure states and no {\it a priori} information
in the ancilla, the right-hand side of the last inequality is the
well-known Helstrom bound \cite{helstrom}. It is the probability of
correctly distinguishing between two pure states $|s_1^{\otimes
N}\rangle$ and $|s_2^{\otimes N}\rangle$ by optimal strategy.

In the case of pure states, the limit by Theorem 1 is least: the
quantum network that attains this limit was presented in Ref.
\cite{chefles}. So, it is natural to ask whether the transformation
given in Ref. \cite{chefles} reaches the limit in the case of mixed
states (and no {\it a priori} information in the ancilla). In general,
the answer is negative. Really, the transformation given in Ref.
\cite{chefles} has two significant features: (1) it acts on the
Hilbert space of system AB composed of registers A and B; (2) the
initial state of the extra register B is pure. But in Ref.
\cite{rast2} we showed that these two features do not allow to reach
always the limit in the special case of $1\to2$ cloning of
equiprobable mixed states. This example convinces that for
optimization of mixed-state cloning the above features should not be
presupposed. The proof of Theorem 1 emphasizes the key importance of
unitarity of quantum mechanical transformations and nondecrease of
the fidelity function under the operation of partial trace. These
general properties have allowed us to place the global-fidelity limit
without detailing of cloning transformation. It is virtue of the
proposed us approach. The weakness is that our approach does not
minutely elucidate an attainability of the presented bound. Just as
the uncertainty relation succeeds the commutation relation
and the quantum mechanical rule calculating the mean, the limit by
Theorem 1 succeeds the unitarity restriction and general properties
of the fidelity. But accessibility of this limit in general case of
mixed states is an open question. On the whole, for pure states we
comprehend our abilities in the quantum information processing much
better than for mixed states.

\protect\section{Limit for multi-state set}

We are now ready to state the global-fidelity limit when the
state-set ${\mathfrak{A}}$ contains more states than two ($n>2$). To
simplify the exposition, we assume that there is no {\it a priori}
information about the state to be cloned. Note that in this case we
have
\begin{equation}
\alpha_{jk}=\Delta^{(N)}_{jk} \!\ .
\label{restrict}
\end{equation}
As before, we take that {\it a priori} probabilities are
arbitrary; ones are constrained only by $\,p_1+\cdots+p_n=1\,$.

{\bf Theorem 2} {\it The global fidelity $F_G$ of standard $N\to L$
cloning for set $\,{\mathfrak{A}}=\{\rho_1,\ldots,\rho_n\}\,$
is limited above by value}
$$
\frac{1}{n-1}
\!\sum_{1\leq j<k\leq n}\frac{p_j+p_k}{2}
\biggl\{1+\Bigl[ 1 -
\frac{4p_jp_k}{(p_j+p_k)^2}\sin^2(\Delta^{(L)}_{jk}
-\Delta^{(N)}_{jk}) \Bigr]^{1/2}\biggr\} \!\ .
$$

{\bf Proof} In line with the definition of global fidelity,
\begin{equation}
F_G=\sum_{1\leq j\leq n} p_j\cos^2\!\delta_j \!\ .
\label{fid1}
\end{equation}
Dividing each term of the sum into $n-1$ equal parts and regrouping
these, Eq. (\ref{fid1}) can be rewritten as
\begin{equation}
F_G=\frac{1}{n-1}
\sum_{1\leq j<k\leq n}
\bigl(p_j\cos^2\!\delta_j+p_k\cos^2\!\delta_k \bigr)
\!\ ,
\label{fid2}
\end{equation}
Because $\,p_j/(p_j+p_k)+p_k/(p_j+p_k)=1\,$, we can apply Theorem
1 to each term of the sum of Eq. (\ref{fid2}). By Theorem 1 and Eq.
(\ref{restrict}),
\begin{equation}
\frac{p_j}{p_j+p_k}\cos^2\!\delta_j
+\frac{p_k}{p_j+p_k}\cos^2\!\delta_k \leq
\frac{1}{2}\biggl\{1+\Bigl[ 1 -
\frac{4p_jp_k}{(p_j+p_k)^2}\sin^2(\Delta^{(L)}_{jk}
-\Delta^{(N)}_{jk}) \Bigr]^{1/2}\biggr\} \!\ .
\label{fid3}
\end{equation}
Eqs. (\ref{fid2}) and (\ref{fid3}) provide the statement of Theorem 2.
$\blacksquare$

The bound established by Theorem 2 is straightforward extension of
limit given by Theorem 1. So this bound succeeds some features of the
latter. For example, if two probabilities, say, $p_1$ and $p_2$ are
variable and the rest of parameters is fixed, then the bound by
Theorem 2 is a decreasing function of $p_1p_2$ and increases as these
probabilities differ. If some one probability is close to 1 and other
probabilities are small, then the bound is also close to 1. This
behavior is expected, because single known state can be cloned
perfectly. For equal {\it a priori} probabilities, that is $p_j=1/n$,
the limit by Theorem 2 becomes
\begin{equation}
F_G \leq\frac{1}{2}+\frac{1}{n(n-1)}
\sum_{1\leq j<k\leq n}
\cos(\Delta^{(L)}_{jk}
-\Delta^{(N)}_{jk}) \!\ .
\label{fid4}
\end{equation}
In the special case of pure states, Eq. (\ref{fid4}) gives the limit
obtained in Ref. \cite{guo}. So we have extended the preceding result
to the mixed-state cloning. As the ancilla contains some {\it a
priori} information, the following modifications must be made in the
limit by Theorem 2. If
$\,F(\Upsilon_j,\Upsilon_k)>F(\rho_j^{\otimes M},\rho_k^{\otimes
M})\,$ for the given pair $[jk]$, then $\Delta^{(N)}_{jk}$ should be
replaced by $\alpha_{jk}$. If
$\,F(\Upsilon_j,\Upsilon_k)\leq F(\rho_j^{\otimes M},\rho_k^{\otimes
M})\,$ for the given pair $[jk]$, then the respective term of the sum
should be replaced by $(p_j+p_k)$.

Now we shall address the following question. Is the presented bound
on the global fidelity always achievable? In general, it is not the
case. In order to reach the limit given by Theorem 2 we must provide
the equality in Eq. (\ref{fid3}) for all pairs $[jk]$, where
$\,1\leq j<k\leq n\,$. As the arguments of Appendix B show, the
maximum of the left-hand side of Eq. (\ref{fid3})
(= the equality in Eq. (\ref{fid3})) holds only if
\begin{equation}
\delta_j+\delta_k=\Delta^{(L)}_{jk}-\Delta^{(N)}_{jk} \!\ .
\label{fid5}
\end{equation}
So we have the system of $n(n-1)/2$ equations of kind (\ref{fid5}) on
the $n$ variables $\,\delta_1,\delta_2,\ldots,\delta_n\,$. Except
some special cases, this system is incompatible, because the number
of equations is larger than the number of variables. Thus, the
presented limit is somewhat rough. More rigorous approach is to
maximize the right-hand side of Eq. (\ref{fid1}) in the domain of
$n$-dimensional space that specified by the conditions
$\,0\leq\delta_j\leq\pi/2\,$ and
\begin{equation}
\delta_j+\delta_k\leq\Delta^{(L)}_{jk}-\alpha_{jk} \!\ .
\label{fid6}
\end{equation}
The conditions of kind (\ref{fid6}) arise from the unitarity
restriction and general properties of the fidelity function. Here we
come across a typical problem of nonlinear programming (the simple
case $n=2$ of this problem is examined in Appendix B). For
prescribed values of parameters, the desired maximum can be found by
one of appropriate numerical methods. At the same time, it is
difficult to obtain an exact formula for general case, mainly due to
the complexity of domain boundary. But even if we should find it,
we still would not have a complete solution to the problem
of mixed-state cloning. In fact, it is not necessary that bound given
by such a formula be least. So we have restricted our consideration
to obtaining of the limit by Theorem 2. Rough though this limit is,
it has straightforward formation and allows to estimate how quantum
laws constrain merit of state-dependent cloning. Optimization of
general unitary transformation is independent task that should be
subjected in future investigations. Evidently, it is harder than in
the case of pure states.

\protect\section{Conclusion}

In present work we have obtained some new results on the limits of
state-dependent cloning. This is obtained by use of the unitarity of
quantum mechanical transformations and the general properties of
the fidelity function. The achievement consists in that known bounds
on the global fidelity of pure-state cloning are extended in two
significances. In the first place, the mixed-state cloning is
regarded. In the second place, we take into account that the ancilla
state can contain {\it a priori} information about the state to be
cloned. The dependence of limit on the parameter, that roughly
describes the amount of {\it a priori} information, is considered.
The conclusions made look reasonable and contribute to the stronger
no-cloning theorem. In general, the presented bound cannot always be
attained. We have described the possible approach to improvement of
bound. This approach may be used in numerical investigations.

Replacing the squared modulus of the inner product by the fidelity
function, the known global-fidelity limits of state-dependentm
cloning of pure states are valid to mixed quantum states. This fact
maintains an intuitive belief that use of mixed states hardly adds
anything new to our possibilities in the quantum information
processing. We study the mixed states rather because all the real
devices are inevitably exposed to noise. So the pure states used us
will eventually evolve to mixed states. It is in the nature of
things.

\appendix

\section{An existence of sufficient states}

Let $\chi$ and $\omega$ be density operators on a finite-dimensional
Hilbert space ${\cal{H}}_1$ and let $r$ be real number such that
$\,0\leq r\leq F(\chi,\omega)\,$. We shall show an existence of states
$\Lambda$ and $\Upsilon$ those have $\chi$ and $\omega$ as the reduced
states for subsystem under the constraint
$\,F(\Lambda,\Upsilon)=r\,$. We briefly recall the definition of the
fidelity in terms of purufication \cite{jozsa}. A purification of
$\chi$ is any pure state $|X\rangle$ in any extended Hilbert space
$\,{\cal{H}}_1\otimes{\cal{H}}_2\,$ with the property that
\begin{equation}
\chi={\rm Tr}_{2}
\bigl(|X\rangle\langle X|\bigr)
\!\ .
\end{equation}
In general, the dimension of ${\cal{H}}_2$ must not be smaller than
the dimension of ${\cal{H}}_1$. The fidelity is defined by
\begin{equation}
F(\chi,\omega)
=\max\: \bigl|\langle X|Y \rangle\bigr|^2
\!\ ,
\label{fiddef}
\end{equation}
where the maximum is taken over all purifications $|X\rangle$ and
$|Y\rangle$ of $\chi$ and $\omega$ respectively \cite{jozsa}. Take
the purifications $|X\rangle$ and $|Y\rangle$ those give the maximum
in Eq. (\ref{fiddef}). Then we have
\begin{equation}
F(\chi,\omega)
=\bigl|\langle X|Y \rangle\bigr|^2
\!\ .
\label{fidmax}
\end{equation}
Adding the qubit space
$\,{\cal{H}}_3={\rm span}\{|0\rangle,|1\rangle\}\,$, we consider
$\,|X0\rangle=|X\rangle\otimes|0\rangle\,$
and $\,|Y\theta\rangle=|Y\rangle\otimes
(\cos\theta\,|0\rangle+\sin\theta\,|1\rangle)\,$,
which lie in $\,{\cal{H}}_1\otimes{\cal{H}}_2\otimes{\cal{H}}_3\,$.
It is clear, these states are purifications of $\chi$ and $\omega$
respectively, that is
\begin{align*}
\chi &={\rm Tr}_{23}
\bigl(|X0\rangle\langle X0|\bigr)
\!\ , \\
\omega &={\rm Tr}_{23}
\bigl(|Y\theta\rangle\langle Y\theta|\bigr)
\!\ .
\end{align*}
In the case of pure states, the fidelity function is equal to the
squared modulus of the inner product. So by Eq. (\ref{fidmax})
and definitions of states $\,|X0\rangle\,$ and $\,|Y\theta\rangle\,$
we have
\begin{equation}
F\bigl(|X0\rangle\langle X0|,|Y\theta\rangle\langle Y\theta|\bigr)
=F(\chi,\omega) \cos^2\!\theta
\!\ . \label{xoxo}
\end{equation}
Because $\,0\leq r\leq F(\chi,\omega)\,$, the right-hand side of Eq.
(\ref{xoxo}) can be set as equal to $r$ by choice of $\theta$. If we
now take $\,\Lambda=|X0\rangle\langle X0|\,$ and
$\,\Upsilon=|Y\theta\rangle\langle Y\theta|\,$, then
$\,\chi={\rm Tr}_{23}\Lambda\,$, $\,\omega={\rm Tr}_{23}\Upsilon\,$
and $\,F(\Lambda,\Upsilon)=r\,$ too. Thus, if Eq. (\ref{restr}) is
violated, then there are the ancilla states those are sufficient for
the ideal cloning.

\section{A lemma}

Consider the function
\begin{equation}
f(x,y)=p\cos^2\!x+q\cos^2\!y \!\ ,
\label{a1}
\end{equation}
where $p$ and $q$ are positive numbers such that $\,p+q=1\,$. Let
$\,a\in[0;\pi/2]\,$ be a fixed parameter. The range of variables
is stated by conditions $\,x+y\geq a\,$,
$\,0\leq x\leq\pi/2\,$ and $\,0\leq y\leq\pi/2\,$. This domain $D$ is
a square whose left-lower corner is truncated by line
$\,x+y=a\,$ (see Fig. 1). We find the maximum of $f(x,y)$ in the
domain $D$.

{\bf Lemma} {\it The maximum of the function $f(x,y)$ in the domain
$D$ is equal to}
\begin{equation}
f_{max}=\frac{1}{2}\left\{
1+\sqrt{1-4pq\sin^2\! a}\right\} \!\ .
\label{a2}
\end{equation}

{\bf Proof} Interior to the domain $D$,
$\,\partial f/\partial x\not=0\,$ and
$\,\partial f/\partial y\not=0\,$. Therefore, the maximum is reached
on the boundary $\partial D$. Let us consider the boundary segment on
which $\,x+y=a\,$. Eq. (\ref{a1}) can be rewritten as
$$
f(x,y)=\frac{1}{2}\{
1+\cos(x+y)\cos(x-y)+(q-p)\sin(x+y)\sin(x-y)\} \!\ ,
$$
that was observed by the authors of Ref. \cite{chefles}. On the
mentioned segment we have
\begin{equation}
f(x,y)=\frac{1}{2}\{
1+\cos a \cos(2x-a)+(q-p)\sin a \sin(2x-a)\} \!\ .
\label{a3}
\end{equation}
By the standard procedure, we obtain the requirement
\begin{equation}
\tan(2x-a)=(q-p)\tan a \!\ .
\label{a4}
\end{equation}
Since on the considered segment $x$ ranges between $0$ and $a$,
the inequality $\,-a\leq2x-a\leq a\,$ holds , where
$a\leq\pi/2$. Then $\cos(2x-a)$ is nonnegative and
$\sin(2x-a)$ has the same sign as $(q-p)$. We can now reexpress
the cosine and the sine in terms of the tangent. By these expressions
and Eq. (\ref{a4}), the maximum is equal to the right-hand side of Eq.
(\ref{a2}). For the rest of boundary segments, the corresponding
maximums are trivially obtained. It is easy to check that they do not
exceed the right-hand side of Eq. (\ref{a2}) (for $a<\pi/2$ they
are strictly lesser than one). $\blacksquare$

\newpage

\begin{figure}[t!] 
\vskip -20mm
\centering{\mbox{\epsfig{file=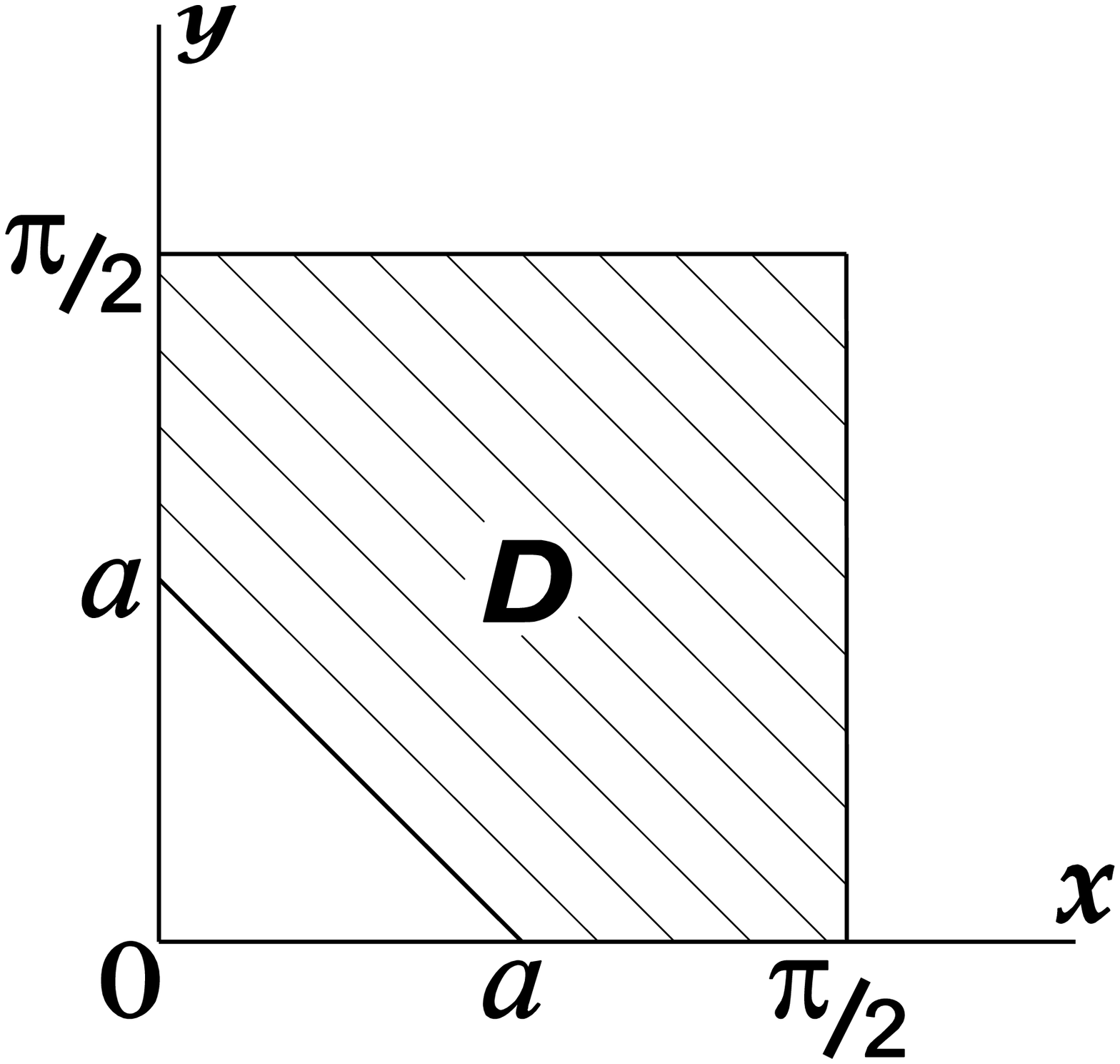,width=11.cm}}}
\vskip 5mm
\caption{The domain $D$, in which we find the maximum of function
         $f(x,y)$, defined by Eq. (\ref{a1}).}
\end{figure}


\begin{references}

\bibitem{wootters}
W.~K.~Wootters and W.~H.~Zurek, Nature {\bf 299}, 802 (1982)

\bibitem{dieks}
D.~Dieks, Phys. Lett. A {\bf 92}, 271 (1982)

\bibitem{barnum}
H.~Barnum, C.~M.~Caves, C.~A.~Fuchs, R.~Jozsa and B.~Schumacher,
Phys. Rev. Lett. {\bf 76}, 2818 (1996)

\bibitem{pati}
A.~Pati and S.~Braunstein, Nature {\bf 404}, 164 (2000)

\bibitem{jozsa1}
R.~Jozsa, e-print quant-ph/0204153

\bibitem{buzek}
V.~Bu\v{z}ek and M.~Hillery, Phys. Rev. A {\bf 54}, 1844 (1996)

\bibitem{hillery}
M.~Hillery and V.~Bu\v{z}ek, Phys. Rev. A {\bf 56}, 1212 (1997)

\bibitem{brass}
D.~Bru{\ss}, D.~P.~DiVincenzo, A.~Ekert, C.~A.~Fuchs, C.~Macchiavello
and J.~A.~Smolin, Phys. Rev. A {\bf 57}, 2368 (1998)

\bibitem{chefles}
A.~Chefles and S.~M.~Barnett, Phys. Rev. A {\bf 60}, 136 (1999)

\bibitem{macchi}
C.~Macchiavello, J. Opt. B: Qantum Semiclassical Opt. {\bf 2}, 144
(2000)

\bibitem{gisin}
N.~Gisin and S.~Massar, Phys. Rev. Lett. {\bf 79}, 2153 (1997)

\bibitem{werner}
R.~F.~Werner, Phys. Rev. A {\bf 58}, 1827 (1998)

\bibitem{keyl}
M.~Keyl and R.~F.~Werner, J. Math. Phys. {\bf 40}, 3283 (1999)

\bibitem{cirac}
J.~I.~Cirac, A.~K.~Ekert and C.~Macchiavello, Phys. Rev. Lett.
{\bf 82}, 4344 (1999)

\bibitem{rast1}
A.~E.~Rastegin, Phys. Rev. A {\bf 66}, 042304 (2002)

\bibitem{rast2}
A.~E.~Rastegin, Phys. Rev. A {\bf 67}, 012305 (2003)

\bibitem{rast3}
A.~E.~Rastegin, e-print quant-ph/0208159

\bibitem{guo}
Y.-J.~Han, Y.-S.~Zhang and G.-C.~Guo, e-print quant-ph/0209094

\bibitem{jozsa}
R.~Jozsa, J. Mod. Optics {\bf 41}, 2315 (1994)

\bibitem{uhlmann}
A.~Uhlmann, Rep. Math. Phys. {\bf 9}, 273 (1976)

\bibitem{zurek}
W.~H.~Zurek, Progr. Theor. Phys. {\bf 89}, 281 (1993)

\bibitem{uhlmann1}
A.~Uhlmann, Rep. Math. Phys. {\bf 45}, 407 (2000)

\bibitem{helstrom}
C.~W.~Helstrom, {\it Quantum Detection and Estimation Theory}
(Academic Press, New York, 1976)

\end{references}
\end{document}